\documentclass[11pt,preprint]{aastex}

\newcommand{\etal}{et~al.\ }
\newcommand{\CIVdblt}{{\rm C}\kern 0.1em{\sc iv}~$\lambda\lambda 1548, 1550$}
\newcommand{\MgIIdblt}{{\rm Mg}\kern 0.1em{\sc ii}~$\lambda\lambda 2796, 2803$}
\newcommand{\NVdblt}{{\rm N}\kern 0.1em{\sc v}~$\lambda\lambda 1238, 1242$}  
\newcommand{\OVIdblt}{{\rm O}\kern 0.1em{\sc vi}~$\lambda\lambda 1031, 1037$} 
\newcommand{\SiIIdblt}{{\rm Si}\kern 0.1em{\sc ii}~$\lambda\lambda 1190, 1193$} 
\newcommand{\SiIVdblt}{{\rm Si}\kern 0.1em{\sc iv}~$\lambda\lambda 1393, 1402$} 
\newcommand{\CII}{\hbox{{\rm C}\kern 0.1em{\sc ii}}}
\newcommand{\CIV}{\hbox{{\rm C}\kern 0.1em{\sc iv}}}
\newcommand{\HI}{\hbox{{\rm H}\kern 0.1em{\sc i}}}
\newcommand{\HII}{\hbox{{\rm H}\kern 0.1em{\sc ii}}}
\newcommand{\Lya}{\hbox{{\rm Ly}\kern 0.1em$\alpha$}}
\newcommand{\Lyb}{\hbox{{\rm Ly}\kern 0.1em$\beta$}}
\newcommand{\FeII}{\hbox{{\rm Fe}\kern 0.1em{\sc ii}}}
\newcommand{\MnII}{\hbox{{\rm Mn}\kern 0.1em{\sc ii}}}
\newcommand{\MgI}{\hbox{{\rm Mg}\kern 0.1em{\sc i}}}
\newcommand{\MgII}{\hbox{{\rm Mg}\kern 0.1em{\sc ii}}}
\newcommand{\NV}{\hbox{{\rm N}\kern 0.1em{\sc v}}}
\newcommand{\OVI}{\hbox{{\rm O}\kern 0.1em{\sc vi}}}
\newcommand{\SiII}{\hbox{{\rm Si}\kern 0.1em{\sc ii}}}
\newcommand{\SiIII}{\hbox{{\rm Si}\kern 0.1em{\sc iii}}}
\newcommand{\SiIV}{\hbox{{\rm Si}\kern 0.1em{\sc iv}}}
\newcommand{\kms}{\hbox{km~s$^{-1}$}}
\newcommand{\cmsq}{\hbox{cm$^{-2}$}}

\shorttitle{{\MgII} Evolution -- $\Omega _{ DLA}$ at $z\simeq 0.05$}
\shortauthors{Churchill}

\begin{document}


\title{~{\MgII} Absorber Number Density at $\lowercase{z} \simeq 0.05$:
Implications for $\Omega_{ DLA}$ Evolution\altaffilmark{1}}

\author{Christopher~W.~Churchill}

\affil{Department of Astronomy and Astrophysics, The Pennsylvania
State University, University Park, PA 16802 \\ {\it
cwc@astro.psu.edu}}

\altaffiltext{1}{Based  in  part  on  observations obtained  with  the
NASA/ESA {\it Hubble Space Telescope},  which is operated by the STScI
for the  Association of Universities for Research  in Astronomy, Inc.,
under NASA contract NAS5--26555}

\begin{abstract}

An unbiased sample of 147 quasar/AGN spectra, obtained with the Faint
Object Spectrograph on board the {\it Hubble Space Telescope\/}, has
been searched for intervening {\MgIIdblt} absorbers over the redshift
range $0 < z \leq 0.15$.  The total redshift path searched is 18.8,
with the survey being $80$\% complete to a $5~\sigma$ rest--frame
equivalent width, $W_{r}(2796)$, of $0.6$~{\AA} and $20$\% complete to
$W_{r}(2796) = 0.3$~{\AA}.  Main results of this work are the
following. (1) Four systems were found, with a mean redshift of
$\left< z \right> = 0.06$, yielding a redshift number density $dN/dz =
0.22^{+0.12}_{-0.09}$ for absorbers with $W_{r}(2796) \geq 0.6$~{\AA}.
This is consistent with the value expected if these systems do not
evolve from higher redshifts ($z=2.2$).  (2) No systems with
$W_{r}(2796)<0.6$~{\AA} were found.  Based upon no evolution
expectations and accounting for the survey completeness, it is a
$2~\sigma$ result to have a null detection of smaller $W_{r}(2796)$
systems.  If this implies a turnover in the low $W_{r}(2796)$ region
of the equivalent width distribution at $z\sim0$, then there is at
least a $25$\% reduction in the average galaxy gas cross section from
$z \leq 0.2$ galaxies.  (3) These systems have strong {\FeII}
absorption and, based upon the results of Rao \& Turnshek (2000, ApJS,
130, 1), are good candidates for damped {\Lya} absorbers (DLAs).  This
translates to a redshift number density of $dN/dz = 0.08
^{+0.09}_{-0.05}$ for DLAs at $z\sim0$.  In tandem with the data
analyzed by Rao \& Turnshek, these results indicate that the redshift
number density of DLAs does {\it not\/} evolve from $z\simeq4$ to
$z\simeq0$.  If the {\HI} mass function does not evolve from
$z\simeq0.5$ to $z\simeq0$, then the cosmological {\HI} mass density,
$\Omega _{ DLA}$, is also deduced to not evolve from $z\simeq4$ to
$z\simeq0$.  These $z\simeq 0$ results for {\MgII}
absorption--selected DLAs are at odds with those based upon 21--cm
emission from {\HI} galaxies by a factor of five to six.

\end{abstract}

%


\section{Introduction}
\label{sec:intro}

At low to intermediate redshift ($0.2 \leq z \leq 1.0$), strong
intervening {\MgIIdblt} absorption systems are known to arise in the
proximity of normal, bright galaxies \citep{bb91,sdp94}.  The projected
size\footnote{Throughout this paper, we assume
$H_0=100$~km~s$^{-1}$~Mpc$^{-1}$.} of absorbing regions around
galaxies was shown to be $R_{\ast} \sim 40$~kpc, with a very weak
proportionality with galaxy $K$ luminosity \citep{steidel95}.
However, there is some evidence that by $z=0$, the {\MgII} absorbing
gas cross sections of galaxies may be less than $30$~kpc
\citep{bbp95}.

\citet{rt2000} showed that strong {\MgII} absorption serves as an
efficient selection method for finding damped {\Lya} absorbers (DLAs;
defined as having $\log N({\HI}) \geq 20.3$ [atoms~{\cmsq}]).  They
found that the equivalent widths of both {\MgII} $\lambda 2796$ and
{\FeII} $\lambda 2600$ are very often greater than $0.5$~{\AA} when
the system is a DLA \citep[also see][]{boisse}.  Measuring the cosmic
evolution in the number densities of DLAs is useful for measuring
evolution in the cosmological mass density of neutral gas, $\Omega
_{ DLA}$.  Selecting DLAs using {\MgII} absorption renders the
computation of $\Omega _{ DLA}$ directly proportional to the
measured redshift number density for {\MgII} systems.  Using this
approach, Rao \& Turnshek have shown that there is no compelling
evidence that DLAs evolve from $z=4$ to $z=0.5$, though currently
published 21--cm emission surveys require strong evolution to $z=0$.
This would suggest a strong evolution in $\Omega _{ DLA}$ at the
lowest redshifts.

A measurement of the redshift number density, $dN/dz$, for {\MgII}
systems at $z<0.2$ could be useful to either corroborate or to
challenge these findings and to place constraints on scenarios of
global galaxy evolution.  If {\MgII} absorbers evolve away over the
interval $0 < z < 0.2$, observed as a down--turning break in the
$dN/dz$ distribution, it would imply that either the number or typical
size of the absorbers is decreasing.  If there is a paucity of large
equivalent width systems having strong {\FeII} absorption, this would
indirectly indicate strong evolution in DLAs.

Measuring $dN/dz$ at $z<0.2$ requires ultraviolet (UV) spectroscopy
(because the {\MgII} doublet is not redshifted to wavelengths redward
of the atmospheric cut off).  There exists a large database of UV
spectra obtained with the Faint Object Spectrograph (FOS) on board the
{\it Hubble Space Telescope\/} ({\it HST\/)}.  Because of the G270H
grating wavelength coverage, especially the upper wavelength cut off,
these data are tailored for comparing the $dN/dz$ of {\MgII} absorbers
from $z=0$ to $z=0.15$ to that published by Steidel \& Sargent
(1992, hereafter SS92\nocite{ss92}) for $0.2 \leq z \leq 2.2$.  They
observed 103 quasars, covering a total redshift path of $\Delta Z
\simeq 130$.  Fortuitously, the FOS spectra have the same resolution
and similar pixel sampling rate as the SS92 spectra.  A smaller, lower
resolution, {\MgII} survey by \citet{boisse92} covered redshifts
$0.2$--$0.4$, with very non--uniform sensitivity down to $z=0.1$.
They observed a total of 35 quasars, covering a total redshift path of
$\Delta Z \simeq 8$.

In this paper, the redshift number density of {\MgII} absorbers with
$z<0.15$ is measured using 147 archival FOS/{\it HST} spectra with a
total redshift path of $\Delta Z \simeq 19$.  In \S~\ref{sec:data},
the data, the quasar and active galactic nuclei (AGN) selection, and
the data analysis are presented.  Results are given in
\S~\ref{sec:results} and discussed in the context of the {\MgII}
absorber redshift number density measured for $0.2 \leq z \leq 2.2$.
In \S~\ref{sec:discuss}, the implications for evolution in $\Omega
_{ DLA}$ are discussed.  A brief discussion of possible evolution
in the gas cross sections of galaxies is presented in
\S~\ref{sec:Xsec}.  Concluding remarks are given in
\S~\ref{sec:conclude}.

%
%

\section{Data and Analysis}
\label{sec:data}

A total of 207 FOS/{\it HST\/} quasar/AGN G270H spectra have been
archived.  The G270H spectra cover the wavelength region $2225$ to
$3280$~{\AA} and provide a redshift coverage for {\MgIIdblt}
absorption from $z\simeq0$ (Galactic) to $z=0.15$.  These spectra have
a resolution of $R= 1300$ with four diodes per resolution element.

Many of the FOS spectra were observed and published by the {\it HST\/}
Quasar Absorption Line Key Project \citep[KP,][]{cat1,cat2,cat3} and
have kindly been made available for this work.  The remaining archived
spectra were generously provided by Dr.\ S. Kirhakos in fully reduced
form and with continuum normalization.  These reductions were
performed using KP software \citep{dpsKP} in the same fashion that
Dr.\ Kirhakos reduced the KP sample; thus, there has been uniform
treatment of the reductions of all spectra for the {\MgII} survey.

Quasars and/or AGN with emission redshifts larger than 1.7 were
removed from the database.  This constraint was applied to ensure that
the quasar {\Lya} emission line was {\it not\/} redward of the
wavelength region covered by the G270H spectra.  This was to avoid
confusion with the {\Lya} forest and/or Lyman series lines of
relatively strong, higher redshift metal--line absorbers.

For each quasar/AGN in the resulting sample, the search for {\MgII}
absorption was limited to wavelength regions 1000~{\kms} redward of
Galactic {\MgII} absorption in order to avoid possible confusion with
high velocity cloud and/or Galactic absorption.  The search was also
limited to redward of the {\Lya} emission line peak. Again, this was
to avoid confusion with {\Lya} forest and/or Lyman series lines.
Since this survey is for {\it intervening\/} {\MgII} absorption, the
search was also limited to $5000$~{\kms} blueward of the quasar/AGN
{\MgII} emission line.  Application of this constraint hopefully
avoided the possible inclusion of any absorption that may be intrinsic
to the quasars themselves.

Of the original 207 spectra, 60 were eliminated due to the above
selection criteria.  Of the 147 spectra remaining, 104 had full G270H
wavelength coverage (1000~{\kms} redward of Galactic {\MgII})
available for the survey.  The partial wavelength regions surveyed in
the remaining 43 spectra ranged from a small $\sim 40$~{\AA} to just
short of $\sim 440$~{\AA}, with a typical coverage of
$200$--$300$~{\AA}.

The equivalent width detection threshold in each pixel was computed,
and the identification of absorption features and {\MgII} doublet
candidates was performed using the software of \citet{weak,archiveI}.
Candidate doublets were then examined visually for inclusion into the
list of absorbers.  The equivalent widths were measured using the
interactive fitting software of \citet{archiveI}, which allows
multiple Gaussian fits to resolved lines.  The {\MgII} rest--frame
equivalent widths are presented in Table~1.

\subsection{Accounting for Biased Lines of Sight}

The KP spectra represent an unbiased sample of quasar/AGN lines of
sight, in that the sample was objectively compiled without
consideration of {\it a priori\/} knowledge of absorption systems.
However, the archive spectra do {\it not\/} constitute an unbiased
sample.  Quasars/AGN observed for the purpose of finding metal--line
absorption near low redshift galaxies would introduce a bias into this
survey.  Inclusion of {\MgII} systems from biased lines of sight would
yield a measured $dN/dz$ larger than the true value.  Conversely,
including biased lines of sight along which no systems are found would
result in a larger total redshift path for the survey, and thus yield
a measured $dN/dz$ smaller than the true value.

In the final sample of quasars/AGN, 74 were observed for the Key
Project and are thus unbiased lines of sight (apart from concerns for
gravitational lensing).  The remaining 73 spectra are a heterogenous
sample from the point of view of why the quasars/AGN were observed.  A
study of the on--line proposal abstracts for these quasars revealed
that, for all but five, the scientific motivations for the
observations were either to study the {\Lya} forest lines or the
environment surrounding or intrinsic to the quasar/AGN themselves.

The five biased lines of sight were excluded from the survey
statistics and all subsequent analysis.  However, they were included
in the survey for systems in order to ``test'' the search algorithm.
The previously published systems that lie in the redshift range
searched in this survey were found and are discussed below.  


\section{Results}
\label{sec:results}

The automated doublet finding software yielded a total of 16 {\it
candidate\/} {\MgII} doublets.  Eight were clearly blends of
known metal lines from other, known intervening systems.  One of the
remaining eight candidates ($z=0.0929$ in PG $1634+706$) was an
artifact of flat fielding residuals \citep[e.g.,][]{jannuzi-hartig}; its
redshifted $\lambda 2796$ and $\lambda 2803$ transitions were
misaligned by $4~\sigma$.  A second, large equivalent width candidate
($z=0.0576$ in B2 $1028+313$) was dropped from the sample because it
was a spurious feature; the absorption had no doublet structure and
there were no corroborating {\FeII} transitions (which are common for
very strong systems).

The remaining six candidates were found to be bonafied {\MgII}
systems, particularly when {\FeII} $\lambda 2600$ and/or {\MgI}
$\lambda 2852$ absorption corroborated the {\MgII} absorption.
Four of these systems were found along unbiased lines of sight,
whereas two were found along biased lines of sight.  These two lines
of sight and the two systems were dropped from the survey analysis,
as described below.

\subsection{The Four Unbiased Systems}

\subsubsection{$3$C~$232$; $z_{em}=0.533$; $z_{abs}=0.0050$}
\begin{figure}[p]
\plotone{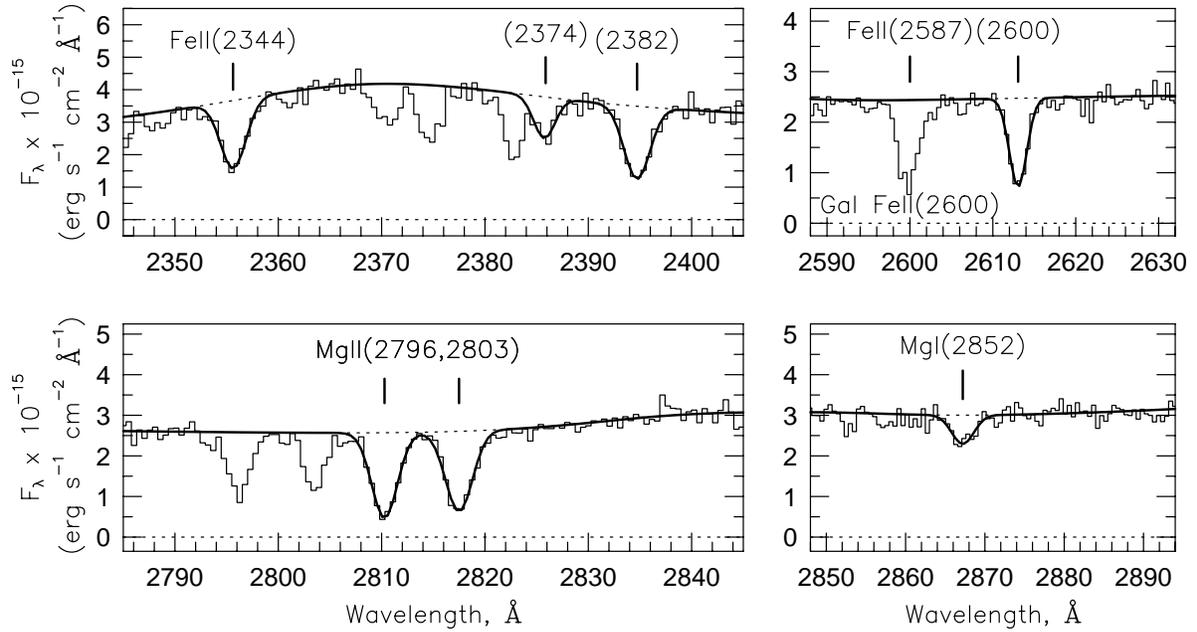}
\caption{The detected transitions for the $z=0.0050$ system toward
3C~232.  The identifications are given above each line. Ticks above
the continuum mark the line centers of the Gaussian fits.  The solid
curves through the data are the Gaussian fits used for measuring the
equivalent widths.  Note that {\FeII} $\lambda 2374$ is partially
blended with Galactic {\FeII} $\lambda 2382$.  Also note that {\FeII}
$\lambda 2587$ (not fit) is blended with Galactic {\FeII} $\lambda
2600$.
\label{fig:3c232}}
\end{figure}

This object was observed for the Key Project, and is thus an unbiased
line of sight.  A subsequent literature search revealed that this
system was cataloged by \citet{cat3}.  Both a G130H and G190H spectrum
were available.  The G130H covered the location of {\Lya} absorption
for this system, but it was heavily obscured by Galactic geocoronal
{\Lya} emission.  The signal strength was too low to discern whether
{\SiIVdblt} or {\CIVdblt} absorption was present.  {\FeII} $\lambda
2344$, 2374, 2382, and 2600 absorption and {\MgI} $\lambda 2852$
absorption were measured.  The data are presented in
Figure~\ref{fig:3c232} and the rest--frame equivalent widths are
presented in Table~2.

\subsubsection{PKS~$0439-433$; $z_{em}=0.593$; $z_{abs}=0.1012$}
\begin{figure}[p]
\plotone{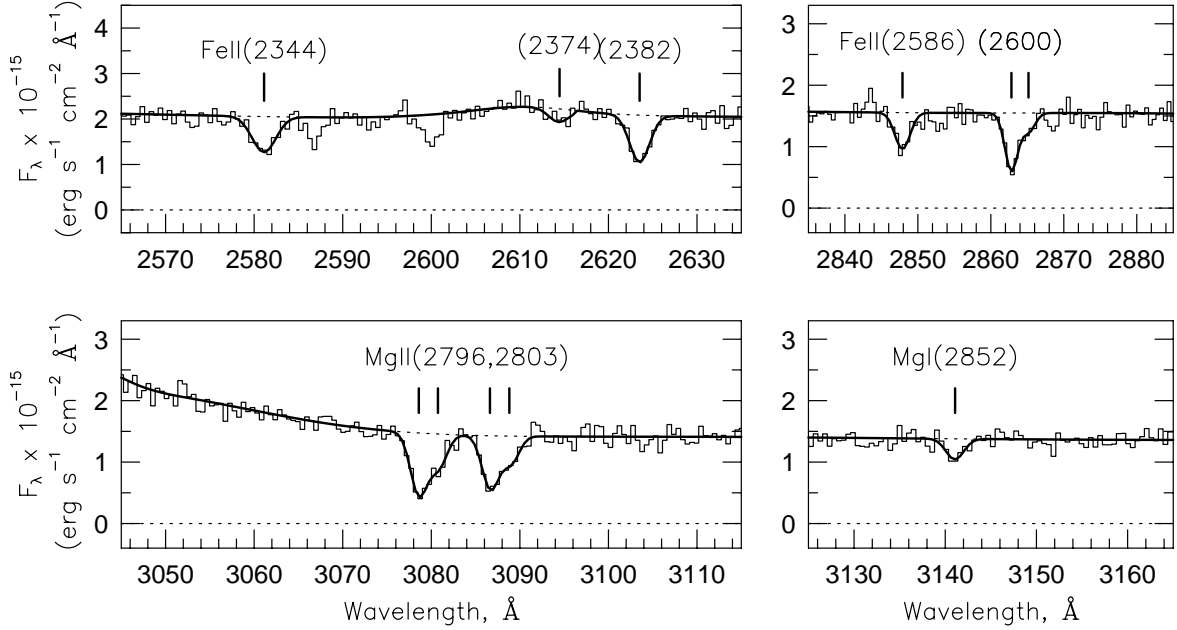}
\caption{The same as Figure~\ref{fig:3c232}, but for the $z=0.1012$
system toward PKS~$0439-433$.  
\label{fig:pks0439}}
\end{figure}

This object was also observed for the Key Project, and is thus an
unbiased line of sight.  As with the 3C~232 absorber, a subsequent
literature search revealed that this system was cataloged by
\citet{cat3}.  The {\MgII} absorption profiles are resolved and
required a two--component Gaussian fit.  A G190H spectrum was also
available for this line of sight.  {\FeII} absorption is strong in
this system.  {\MgI} $\lambda 2852$ was also detected.  The data are
presented in Figure~\ref{fig:pks0439} and the rest--frame equivalent
widths are presented in Table~2.

\subsubsection{Q~$1327-206$; $z_{em}=1.169$; $z_{abs}=0.0174$}
\begin{figure}[p]
\plotone{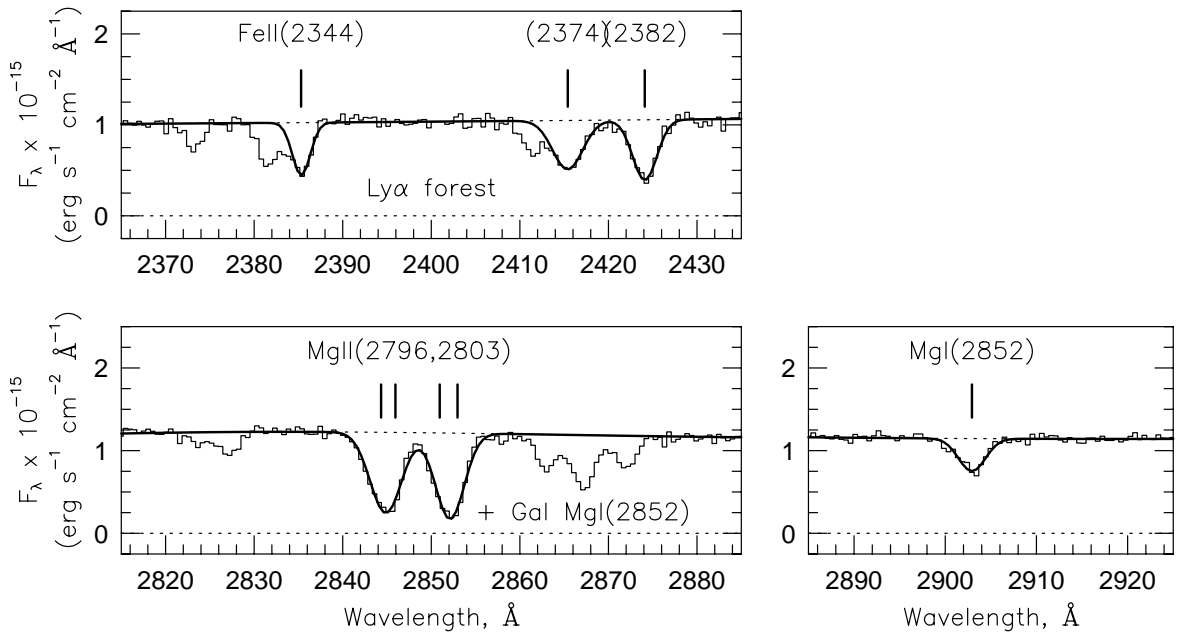}
\caption{The same as Figure~\ref{fig:3c232}, but for the $z=0.0174$
system toward Q~$1327-206$. Note that the {\FeII} lines are in the
{\Lya} forest, and there may be some blending effects.  Also note that
{\MgII} $\lambda 2803$ is blended with Galactic {\MgI} $\lambda 2852$.
\label{fig:q1327}}
\end{figure}

This object was observed for a program (PID 5654) to catalog {\Lya}
forest absorptions lines; thus, the line of sight is unbiased for
intervening--metal line absorbers.  The stronger absorption profiles
are resolved and required a two--component Gaussian fit.  {\MgII}
$\lambda 2803$ is blended with Galactic {\MgI} $\lambda 2852$; this
resulted in an unphysical doublet ratio for the red component of the
Gaussian fit.  Only a G270H spectrum was available for this line of
sight.  Because of the relatively high redshift of this emission line
object, the strong {\FeII} transitions fell in the {\Lya} forest.
{\FeII} $\lambda 2600$ is just redward of the {\Lya} emission, but it
is hopelessly blended in a strong absorption complex.  {\MgI} $\lambda
2852$ is quite strong.  The data are presented in
Figure~\ref{fig:q1327} and the rest--frame equivalent widths are
presented in Table~2.

\subsubsection{PG~$1427+480$; $z_{em}=0.221$; $z_{abs}=0.1203$}
\label{sec:nonDLA}
\begin{figure}[p]
\plotone{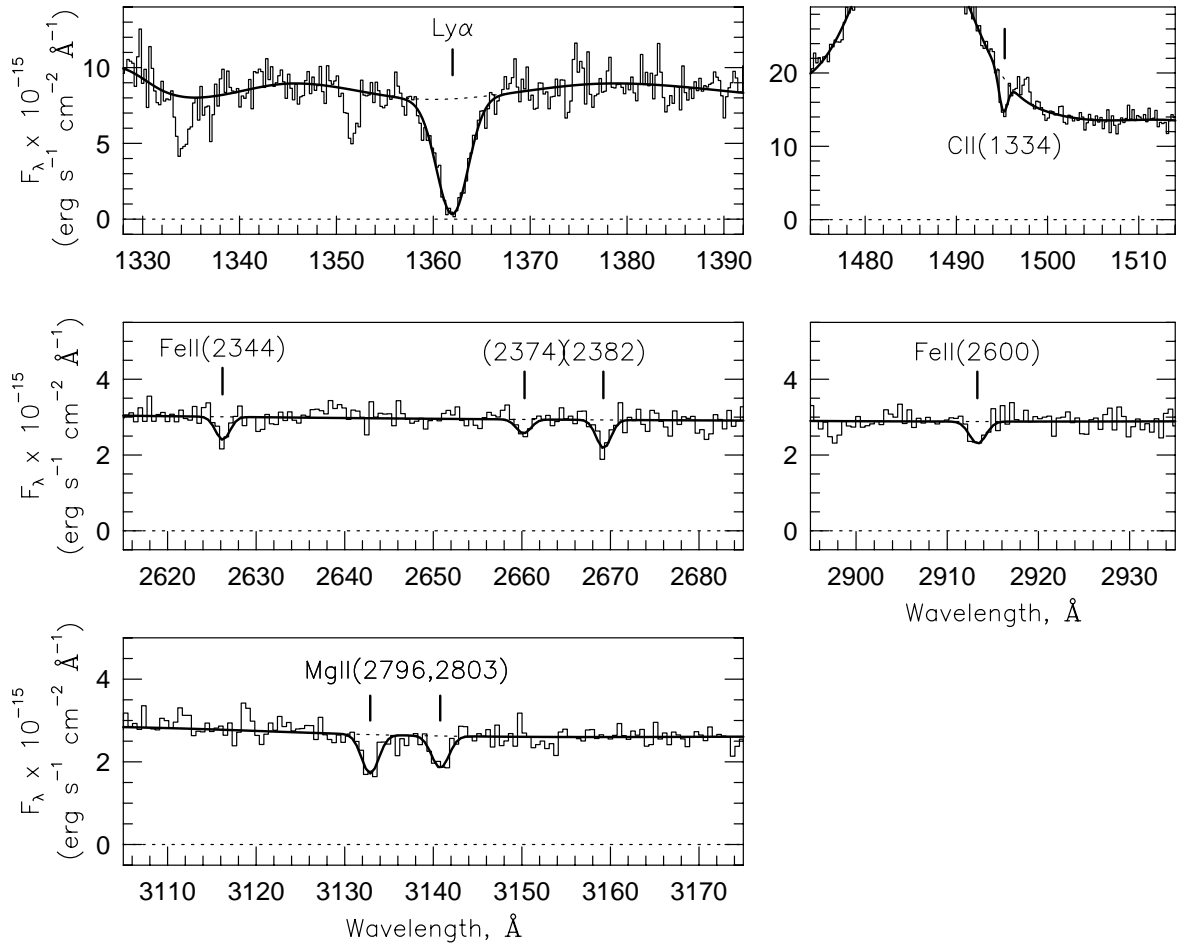}
\caption{The same as Figure~\ref{fig:3c232}, but for the $z=0.1203$
system toward PG~$1427+480$.
\label{fig:pg1427}}
\end{figure}

This object was observed for a program (PID 6781) to investigate the
ionizing continuum in AGN; this is an unbiased line of sight.  The
{\MgII} system is the weakest of the four and the only one in which
the profiles are unresolved at FOS resolution (subsequently, the
widths of the Gaussians used for the equivalent width measurements
were held constant at the value of the FOS instrumental spread
function; this leaves a somewhat significant residual to the fit in
the line cores). Both a G130H and a G190H spectrum were available for
this line of sight.  The {\Lya} rest--frame equivalent width is
$3.95\pm0.26$~{\AA}, indicating that this systems is not a DLA.  No
{\SiIVdblt} nor {\CIVdblt} absorption was detected to a $5~\sigma$
detection limit of $0.27$~{\AA} and $0.36$~{\AA}, respectively.
{\CII} $\lambda 1334$ was detected.  Also detected were {\FeII}
$\lambda 2344$, 2374, 2382, and 2600.  {\MgI} $\lambda 2852$ was not
detected to 0.49~{\AA} ($5~\sigma$).  The data are presented in
Figure~\ref{fig:pg1427} and the rest--frame equivalent widths are
presented in Table~2.

\subsection{The Two Biased Systems}

\subsubsection{PKS~$0454+039$; $z_{em}=1.345$; $z_{abs}=0.072$}
In the spectrum of PKS $0454+039$ \citep{boisse}, {\MgII} absorption
at $z=0.072$ is associated with a post--star burst dwarf galaxy
\citep{ccs0454}.  Boiss\'{e} \etal state that the presence of this
galaxy was ``one additional motivation for observing this quasar''.
Thus, this quasar was dropped from this survey.

\subsubsection{Q~$1219+047$ $z_{em}=0.094$; $z_{abs}=0.0052$}
\citet{bbp96} published this system,
which arises in the inclined galaxy M61.  This quasar was
observed because it was a sightline that ``lies fortuitously behind [a
galaxy] whose existence is already known.''  Thus, this line of sight
was also dropped from this survey.

\subsection{Redshift Path Density}

\begin{figure}[p]
\plotone{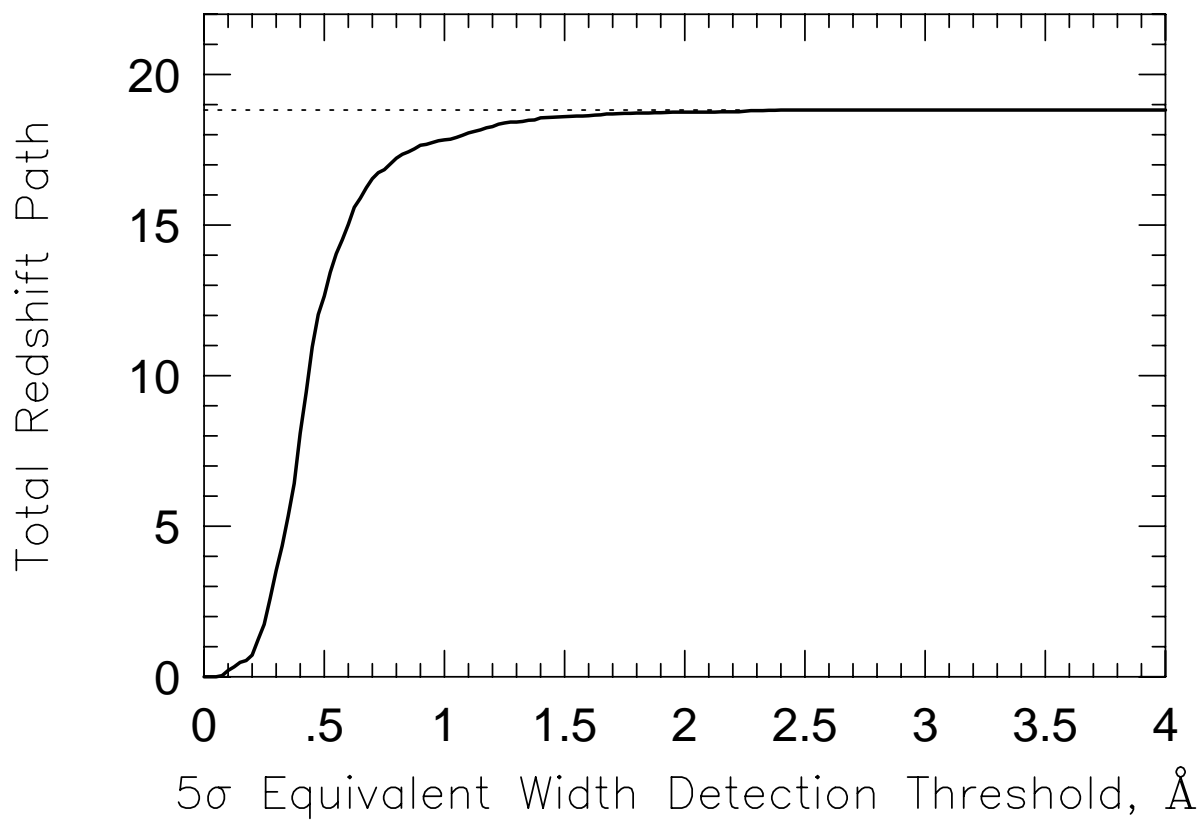}
\caption{The total redshift path density, $g(W_{r})$, as a function of
the $5~\sigma$ equivalent width detection threshold.  The dotted line
at $g(W_{r}) = 18.82$ is the maximum total redshift path density,
which is achieved for $W_{r} > 2.4$~{\AA}.
\label{fig:gwz}}
\end{figure}

The total redshift path, over which an absorber with a specified
rest--frame equivalent width, $W_{r},$ could be detected in an
unbiased survey was computed using the the formalism of \citet{ltw87},
(their Eqs.\ 6 \& 7) and SS92\nocite{ss92} (their Eqs.\ 1 \& 2).  The
total redshift path density, $g(W_{r})$, of the survey is shown in
Figure~\ref{fig:gwz}, where we have adopted the notation of
\citet{ltw87} and a $5~\sigma$ detection limit.  The total redshift
path is $18.82$ above $W_{r}(2796) \simeq 1$~{\AA} and then drops
rapidly below $\sim 0.5$~{\AA}.  The survey is $80$\% complete at
$W_{r}(2796) = 0.6$~{\AA} and $20$\% complete at $W_{r}(2796) =
0.3$~{\AA}.

The redshift number density, $dN/dz,$ is given by $\sum_{i} 
[g_{i}(W_{r})]^{-1}$ and its variance by $\sum_{i} 
[g_{i}(W_{r})]^{-2}$ \citep[Eqs.\ 9 \& 10 of][]{ltw87}.
The sum is over all {\MgII} absorbers having $W_{r}(2796) \geq
W_{min}$, where $W_{min}$ is a chosen lower cut off equivalent width.
The measured number densities are,
\begin{equation}
dN/dz = 0.22 ^{+0.12}_{-0.09} \quad {\rm for} \quad 
W_{min} = 0.6~{\rm \AA},
\end{equation}
and
\begin{equation}
dN/dz = 0.16 ^{+0.09}_{-0.05} \quad {\rm for} \quad 
W_{min} = 1.0~{\rm \AA},
\end{equation}
with mean redshifts $\left< z \right> = 0.06$ and $\left< z \right> =
0.04$, respectively.

\subsection{Evolution}

The redshift evolution of $dN/dz$ can be written
\begin{equation}
dN/dz = N_0(1+z)^{\gamma} , 
\label{eq:dndz}
\end{equation}
where $N_0$ gives the $z=0$ expectation and the power law index,
$\gamma$, parameterizes the redshift evolution.  A non--evolving
population of objects would have $0.5 \leq \gamma \leq 1.0$ for $0.5
\geq q_0 \geq 0$, respectively.  SS92 divided their large, higher
redshift sample into three subsamples; MG1 ($W_{min}=0.3$~{\AA}), MG2
($W_{min}=0.6$~{\AA}), and MG3 ($W_{min}=1.0$~{\AA}).  

For subsample MG1, SS92\nocite{ss92} found $dN/dz=0.97 \pm 0.10$ with
$\gamma = 0.78 \pm 0.42$ at $\left< z \right> = 1.12$.  The smallest
equivalent width detected in this survey is $W_{r}(2796) =
0.78$~{\AA}; no systems with $0.3 \leq W_{r}(2796) < 0.6$~{\AA} were
observed.  In this equivalent width bin, extrapolation of SS92 data
yield $dN/dz = 0.31^{+0.22}_{-0.16}$ at $z=0.04$; a null detection in
this {\it bin\/} is a $\simeq 2~\sigma$ event.  This is suggestive of
a paucity of smaller equivalent width systems at very low redshift,
which would imply a turnover in the {\MgII} equivalent width
distribution compared to intermediate and higher redshifts.

The expected number of systems with $W_{min}=0.3${\AA} from this
survey can be estimated using the {\MgII} equivalent width
distribution and correcting for the survey completeness.
At $\left< z \right> \simeq
0.7$, this distribution was observed to follow a power law, $f(W_{r}) =
CW_{r}^{-1}$, down to $W_{r}(2796) \geq 0.02$~{\AA} \citep{weak}.  The
normalization constant, $C$, is defined so that $f(W_{r})$ is the
number of systems with equivalent width $W_{r}$ per unit equivalent
width per unit redshift at $\left< z \right>$.  This is written,
\begin{equation}
\int _{W_{min}}^{W_{max}} g^{\prime}(w)f(w)dw = \frac{dN}{dz}(\left< z
\right> ) ,
\end{equation}
where the integral is performed over the minimum and maximum
$W_{r}(2796)$ found in the survey, and $g^{\prime}(W_{r})$ is the
completeness of the survey at $W_{r}$.

Assuming $f(W_{r}) = CW_{r}^{-1}$ and normalizing to the $W_{min} =
0.6$ subsample at $\left< z \right> = 0.04$, the expected $dN/dz$ for
$W_{min} = 0.3$~{\AA} is $0.32 ^{+0.12}_{-0.09}$.  The quoted errors
do not include the uncertainty in the power--law slope ($= 10$\%) of
$f(W_{r})$.  

\begin{figure}[p]
\plotone{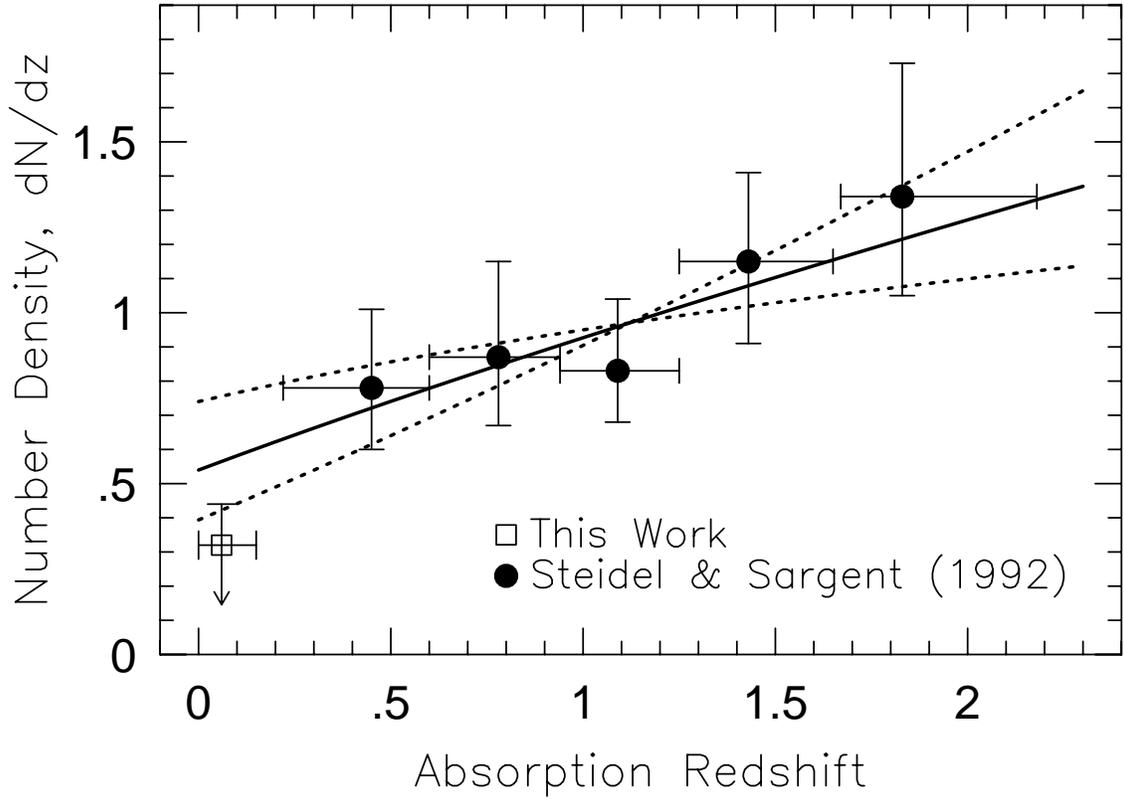}
\caption{Redshift number density vs.\ absorption redshift for
$W_{r}(2796) \geq 0.3$~{\AA}.  The open square datum point, $(\left<
z\right> =0.06,dN/dz \leq 0.32)$, is the value from the FOS
spectra over the redshift range $0 \leq z \leq 0.15$, assuming no
evolution in the equivalent width distribution (see text).  The
filled circular data points are taken from SS92 for $0.2 \leq z \leq 2.2$.
The solid curve through the higher redshift data is their best fit to
Eq.~\ref{eq:dndz}.  The dashed curves show the $1~\sigma$ slopes.
\label{fig:dndz}}
\end{figure}

In Figure~\ref{fig:dndz}, the SS92 data (filled circles) for subsample
MG1 and the best fit (solid curve) are shown.  The dashed curves show
how the slope varies with the $1~\sigma$ uncertainty in power law.
For subsample MG1, the SS92 results extrapolate to give $dN/dz =
0.56^{+0.19}_{-0.14}$ at $\left< z \right> =0.06$.  If the true
equivalent width distribution declines for small $W_{r}(2796)$, then
the expected value of $dN/dz$ for $W_{min}=0.3$~{\AA} serves as an
upper limit compared to a non--evolving equivalent width distribution.
The expected $dN/dz$ for $W_{min}=0.3$~{\AA} from this survey is
plotted as a open--square data point.

Accounting for the uncertainty in the expected $dN/dz$, it is
consistent with the SS92 extrapolated value, {\it assuming a
non--evolving equivalent width distribution}.  This consistency lends
further support to the suggested turnover for $W_{r}(2796) <
0.6$~{\AA}.  However, one cannot place a significance level on this
statement without {\it a priori\/} knowledge of the equivalent width
distribution. Again, based only upon the observed data, the lack of a
detection in the bin covering $0.3 \leq W_{r}(2796) < 0.6$~{\AA} is
significant to the $2~\sigma$ level.  As such, the data are suggestive
of a turnover in the equivalent width distribution for $W_{r}(2796) <
0.6$~{\AA}.

\begin{figure}[p]
\plotone{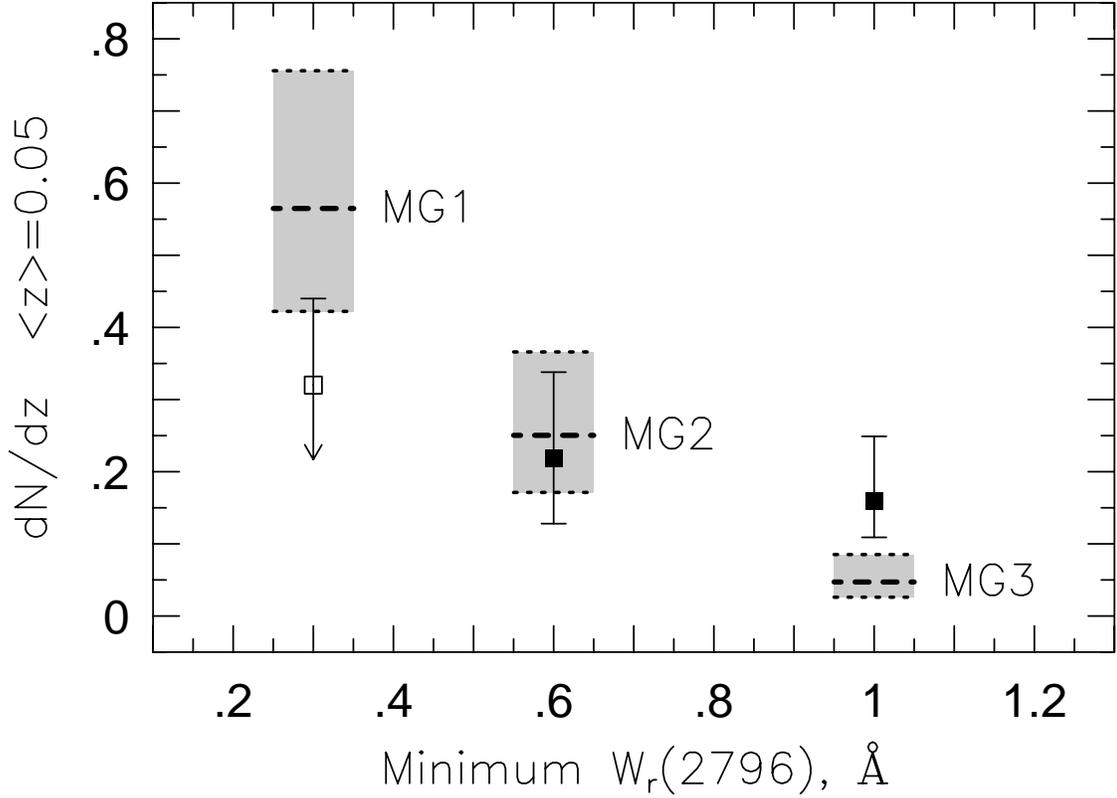}
\caption{The redshift number density at $\left< z \right> \simeq 0.05$
for three minimum equivalent width cut offs.  The shaded regions give
the expected values based upon the extrapolated fits to the higher
redshift data of SS92 for their three subsamples, MG1, MG2, and MG3
(corresponding to $W_{min}= 0.3$, $0.6$, and $1.0$~{\AA},
respectively).  The thick long--dash lines are the best values and the
short--dash lines are the $1~\sigma$ uncertainties.  The solid data
points are the values measure in this work. The open datum point is
the ``measured'' upper limit for $W_{min}= 0.3$~{\AA}, assuming no
evolution in the equivalent width distribution (see text)
\label{fig:lowz}}
\end{figure}

The extrapolation for the SS92 subsample MG2 at $\left< z \right>
=0.06$ is $dN/dz = 0.25^{+0.11}_{-0.08}$ and for subsample MG3 is
$dN/dz = 0.05^{+0.04}_{-0.02}$ at $\left< z \right> =0.04$.  These
mean redshifts are taken to match those of the $W_{min} = 0.3$ and
$0.6$~{\AA} subsamples of this survey.  In Figure~\ref{fig:lowz},
these extrapolated low redshift $dN/dz$ and their $1~\sigma$ ranges
are plotted as horizontal lines in shaded regions for each of the SS92
subsamples.  The measured $dN/dz$ for each $W_{min}$ are also plotted
in Figure~\ref{fig:lowz}, where the ``no--evolution predicted'' value
(upper limit) is plotted for $W_{min}=0.3$~{\AA}.  The error bars are
$1~\sigma$.  These data are also listed in Table~3.

The measured redshift number density for $W_{min}=0.6$~{\AA} is fully
consistent with the extrapolations from higher redshift.  This would
suggest that there is no departure from the SS92 results all the way
to $z\sim0$ for subsample MG2.  For $W_{min} = 1.0$~{\AA}, there is a
marginal suggestion for an overabundance of very strong {\MgII}
systems at $z\sim0$.  However, this is not a significant result
($1.3~\sigma$).

With regard to the equivalent width distribution, the $\left< z
\right> = 0.05$ results are slightly at odds with the $\left< z
\right> = 0.25$ results of \citet{boisse92}.  They found five systems
at redshifts 0.1920, 0.2216, 0.2220, 0.2255, and 0.3937 with
rest--frame equivalent widths 0.33, 0.52, 0.55, 0.51, and 0.34~{\AA},
respectively.  Their redshift number density is in agreement with SS92
for $W_{min} = 0.3$~{\AA}, but is marginally lower for $W_{min} =
0.6$~{\AA}.  If there is a turnover at very low redshift in the
equivalent width distribution below $W_{r}(2796)=0.6$~{\AA}, the SS92
and Boiss\'{e} \etal data would require that it occur rapidly from
$z\simeq 0.2$ to $z \simeq 0$.


\section{Implications for DLAs at $\lowercase{z}\simeq 0$}
\label{sec:discuss}

\subsection{Redshift Number Density}

Based upon an efficient {\MgII}--selection method for finding and
counting damped {\Lya} absorbers (DLAs), with $\log N({\HI}) \geq
20.3$ [atoms {\cmsq}], \citet{rt2000} showed that there is little to
no indication that the redshift number density of DLAs evolves from
$z\simeq4$ to $z\simeq0.5$.  However, at $z=0$, the inferred redshift number
density of {\HI} galaxies with $\log N({\HI}) \geq 20.3$
[atoms~{\cmsq}], based upon 21--cm emission surveys, is apparently a
factor of five to six lower \citep{rb93,rtb95,zwaan97,zwaan99}.  This
implies strong evolution of $dN/dz$ for DLAs, though it is not clear
if it is rapid from $z\simeq0.5$ or more gradual from $ z\simeq 1.5$
\citep{rt2000}.

Rao \& Turnshek found that half of all {\MgII} systems with
$W_{r}({\MgII} ~\lambda 2796)$ and $W_{r}({\FeII}~\lambda 2600)$ both
greater than $0.5$~{\AA} are DLAs with $\log N({\HI}) \geq 20.3$
[atoms {\cmsq}].  All four of the {\MgII} absorbers found in this
survey are members of the $W_{min}=0.6$~{\AA} subsample and have
strong {\FeII} $\lambda 2600$ absorption; they are good candidates for
DLAs.  Statistically, if half are DLAs, then the implied redshift
number density of DLAs at $z\sim0$ is $dN/dz \simeq
0.08^{+0.09}_{-0.05}$, which is fully consistent with the values found
by Rao \& Turnshek at intermediate redshifts (for $W_{min} =
0.6$~{\AA}).  Note that the $z=0.1203$ system is {\it not\/} a DLA
(see \S~\ref{sec:nonDLA}).  However, eliminating this system and using
the $W_{min} = 1.0$~{\AA} subsample yields the same value for $dN/dz$.

From Figure~29 of \citet{rt2000}, we see that the $dN/dz$ at $ \left<
z \right> \simeq 0.05$ from this work is consistent with their
measured $dN/dz$ at $ \left< z \right> \simeq 0.5$ and $ \left< z
\right> \simeq 1.15$.  Thus, a main result of this {\MgII} survey is
that there is apparently little to no evolution of DLAs over the full
redshift range $z\simeq 4$ to $z\simeq 0$.  This also places the
previously inferred DLA redshift number density of 21--cm selected,
{\HI} galaxies with $\log N({\HI}) \geq 20.3$ [atoms~{\cmsq}] at odds
with that determined for {\MgII}--absorption selected DLAs at $z\simeq
0$.

The bulk of the local {\HI} resides in the massive--{\HI} spiral
galaxies \citep{zwaan99}.  However, for very low to intermediate
redshifts, DLAs are associated with galaxies having a variety of
morphologies and luminosities ranging down to $\sim 0.1~L_{\ast}$
\citep{lebrun97,rt98,turnshek01}, including low surface brightness
(LSB) galaxies \citep{turnshek01,btj2001}.  In one case a DLA has no
measurable optical counterpart \citep{3c336} nor narrow band H$\alpha$
emission \citep{bouche}.  Unless the integral number of LSB and low
luminosity {\HI}--selected galaxies with $\log N({\HI}) \geq 20.3$
[atoms~{\cmsq}] dominate the number counts (requires a steep
faint--end slope to the {\HI}--selected galaxy luminosity function),
these observational facts remain unreconciled, as do the redshift
number densities.  Some recent 21--cm results are suggestive that the
integral number of LSB and low luminosity {\HI}--selected galaxies may
bring the observations into better agreement
\citep{briggs-privcomm,zwaan-rf}.

The roughly constant redshift density of DLAs from $0\leq z \leq 1$
strengthens arguments based upon $z>1$ data that DLAs are not tracing
galaxies undergoing the bulk of cosmic star formation
\citep[e.g.,][]{rt2000,pettini99,steidel99}, especially since the
global star formation rate evolves (decreases) rapidly below $z=1$.

\subsection{$\Omega_{ DLA}$}

The cosmological {\HI} mass density, $\Omega _{ DLA}$, is
proportional to the product of the redshift number density and the
mean {\HI} column density of the sample of systems.  \citet{rt2000}
found that $\Omega_{ DLA}({\rm ALs})$ from absorption lines (ALs) at $0.5
\leq z \leq 1.5$ is dominated by the largest $N({\HI})$ systems, of
which there is a relatively large fraction in their sample.  That is,
the {\HI} column density distribution function of DLAs is weighted
more heavily toward larger column densities than it is for
{\HI}--selected galaxies.

If the distribution of $N({\HI})$ does not evolve from that observed
by \citet{rt2000} for $\left< z \right> = 0.78$, then the measured
$dN/dz$ from this survey results in an $\Omega_{ DLA}({\rm ALs})$ at
$z\simeq0$ comparable to that found at intermediate redshift.  This is
at odds with the cosmological {\HI} mass density from {\HI}--selected
galaxies, $\Omega_{ DLA}({\rm 21cm})$, at $z\simeq 0$.  This quantity is
a factor of five to six smaller than the value deduced from this work
using absorption line statistics.

An accurate measurement of $\Omega _{ DLA}({\rm 21cm})$ is
dependent upon the details of the faint--end slope of the 21--cm
emission, {\HI} mass function.  The bulk of the local {\HI} resides in
the massive--{\HI} spiral galaxies \citep{zwaan99}, so that this slope
would need to be quite steep if {\HI}--rich dwarf and LSB galaxies
were to significantly contribute to the value of $\Omega _{
DLA}({\rm 21cm})$.  Even if the integral number of low luminosity,
{\HI}--rich dwarf and large cross--section LSB galaxies reconciled the
absorption--selected DLA and {\HI}--selected DLA redshift number
densities at $z\simeq0$, it would not necessarily translate to a
similar reconciliation for $\Omega _{ DLA}({\rm ALs})$ and $\Omega
_{ DLA}({\rm 21cm})$.  This is because the {\HI}--rich dwarf and
LSB galaxies have relatively low {\HI} masses \citep{zwaan-rf}, so
even if their large cross sections reconcile the $dN/dz$ numbers,
their combined mass/numbers are likely not large enough for their
integrated {\HI} mass to raise the value of $\Omega _{ DLA}({\rm
21cm})$ at $z=0$.

As such, it seems unlikely that the factor of five to six discrepancy
between the absorption line data and the 21--cm emission $z \simeq 0$
cosmological {\HI} mass density can be reconciled by integrating to
smaller and smaller decades of {\HI} mass.  

These conclusions are based upon the assumption that there is (1) no
bias for a large mean {\HI} column density from the \citet{rt2000}
{\HI} column density distribution, and (2) no evolution in the {\HI}
column density distribution from $\left< z\right> \simeq 0.8$.  Note
that the distribution at $\left< z\right> \simeq 0.8$ was observed to
significantly evolve from $\left< z \right> \simeq 2.3$ in that the
lower redshift distribution has a greater fraction of large $N({\HI})$
systems \citep{rt2000}.  If this evolutionary trend continued, the
deduced $\Omega _{ DLA}$ would be greater, a result that would
increase the discrepancy between the absorption line and {\HI} data.

\section{Evolution in Galaxy Gas Cross Sections?}
\label{sec:Xsec}

Since $dN/dz$ is proportional to the product $ n\sigma = \pi \Phi
_{\ast} R_{\ast}^{2}$, redshift evolution in $dN/dz$ would imply
redshift evolution in this product.  There is also a dependence on the
faint--end slope of the luminosity function, which has been seen to
weakly evolve from intermediate redshifts \citep[e.g.,][]{marzke98}.
However, this will not be considered here.

For $0.2 \leq z \leq 1$, the normalization of the galaxy luminosity
function, $\Phi _{\ast}$, for {\MgII} absorption selected galaxies
\citep{sdp94} is consistent with that of photometric galaxies surveys
\citep[e.g.,][]{lilly95,ellis96,lin99}.  However, there
is some ambiguity and disagreement as to whether $\Phi_{\ast}$ is
smaller by a factor of $\sim 2$ over the redshift range $0\leq z \leq
0.2$ \citep[e.g.,][]{maddox90,loveday92,bertin97,marzke98}.

Assuming no evolution in $\Phi_{\ast}$, the upper limit on {\MgII}
absorber sizes would be $R_{\ast} \leq \sqrt{0.32/0.56} \cdot 40 =
30$~kpc.  Smaller absorbing cross sections could result from a general
relaxation of ``halo'' gas with time.  As such, this might result in
the average {\MgII} absorber at $z \simeq 0$ having a relatively large
equivalent width and arising at relatively lower impact parameters.
This might explain why no $W_{r}(2796) < 0.6$~{\AA} systems were found
in this survey.  This is also consistent with the results of
\citet{bbp95}, who found $W_{r}(2796) > 0.6$~{\AA} within $10$~kpc for
four out of five galaxies and no absorption down to $0.04 \leq W_{r}
\leq 0.09$~{\AA} ($2~\sigma$) beyond $30$~kpc for nine of ten galaxies.
Unfortunately, they did not sample the impact parameters between
$\sim10$ and $\sim 30$~kpc.

These statements depend upon a relatively small survey in which the
lack of detected small $W_{r}(2796)$ systems could be dominating the
number statistics.  Smaller galaxy gas cross--sections would require
some level of turnover at small $W_{r}(2796)$ in the equivalent width
distribution at $z\simeq 0$.


\section{Conclusion}
\label{sec:conclude}

An unbiased survey for {\MgII} absorption systems at $z < 0.15$ was
conducted in 147 FOS spectra.  The total redshift path was $\Delta z =
18.8$, with a rapid decline below $W_{r}(2796) = 0.5$~{\AA}
($5~\sigma$).   Main results of this work are:

(1) Four ``unbiased'' systems were found, two of which were previously
published \citep{cat3}.  The mean redshift of these systems is
$\left< z \right> = 0.06$.  All have $W_{r}(2796) > 0.6$~{\AA} and
three of the four have $W_{r}(2796) > 2.0$~{\AA}.  Each system has
strong {\FeII} absorption.  They are all candidates for DLAs,
though the smallest equivalent width system ($z=0.1203$ toward PG
$1427+480$) was verified to have $\log N({\HI})$ well below
$20.3$~[atoms {\cmsq}].

(2) The redshift number density of {\MgII} absorbers at $\left< z
\right> = 0.06$ with $W_{min}=0.6$~{\AA} is consistent with the
expected value if these systems do not evolve with respect to the
number density of higher redshift systems.  

(3) No systems in the equivalent width bin $0.3 \leq W_{r}(2796) <
0.6$~{\AA} were found; a null detection in this equivalent width bin
is a $2~\sigma$ result.  This is somewhat suggestive of a turnover in
the equivalent width distribution at $W_{r}(2796) < 0.6$~{\AA}.
If true, this implies that the {\MgII} gas cross sections of galaxies
decreases by at least $\sim 25$\% from $z \geq 0.2$.

(4) Applying the {\MgII} statistics of \citet{rt2000} for selecting
DLAs, it is found that the redshift number density for DLAs at $z =
0.05$ is consistent with the values they measure at $\left< z \right>
=0.5$ and $\left< z \right> =1.15$.  Thus, accounting for very high
redshift data, there is no evidence for evolution in the number
density of DLAs from $z\simeq4$ to $z\simeq0$.

(5) Assuming no evolution in the {\HI} column density distribution
function, the deduced $\Omega _{ DLA}$ at $z\simeq 0$ is also
consistent with the values measure at $\left< z \right> =0.5$ and
$\left< z \right> =1.15$; thus it appears that $\Omega _{ DLA}$
does not evolve from $z\simeq4$ to $z\simeq0$.

(6) At $z\simeq0$, there are now inconsistencies between both the
redshift number density and cosmological mass density of absorption
selected DLAs and those measured from 21--cm emission {\HI}--selected
galaxies.  Even if $dN/dz$ can be reconciled, it is unlikely that
$\Omega _{ DLA}$ will also be reconciled unless the {\HI} column
density distribution function evolves from $z\simeq0.5$ to $z\simeq0$.

With regards to the absorption line data, it must be cautioned that
there are concerns about small numbers statistics (including the
conversion from {\MgII} statistics), biasing due to gravitational
lensing (overestimates $dN/dz$), and the effects of dust in DLAs
(underestimates $dN/dz$).  These issues have been discussed in
\citet{rt2000}.  It would be useful to directly confirm if the strong
{\MgII} absorbers found in this survey are bonafied DLAs.


\acknowledgments

Gratitude is extended to Sofia Kirhakos for providing the entire
FOS/{\it HST\/} archived in fully reduced form.  Thanks also to Buell
Jannuzi and Don Schneider of the {\it HST\/} Quasar Absorption Line
Key Project for insights into the data reduction and FOS instrument.
Frank Briggs, Jane Charlton, Sandhya Rao, David Turnshek, and Martin
Zwaan are thanked for helpful discussions and for comments which led
to an improved manuscript.  This work was inspired by the
presentations and off--line discussions at the ``Rayfest'', held in
honor of Ray Weymann, April 4--6, 2001.


\begin{table}[p]
\begin{center}
\begin{tabular}{rccc}
\multicolumn{4}{c}{TABLE 1. {\sc Unbiased Absorbing Systems}}  \\[0.7ex]\hline\hline
Object         & $z_{abs}$ & $W_{r}(2796)$, {\AA} & $W_{r}(2803)$, {\AA} \\[0.3ex]\hline
3C~232         & $0.0050$  & $2.54\pm0.09$        & $2.45\pm0.10$      \\[0.1ex] 
PKS~$0439-433$ & $0.1012$  & $2.32\pm0.39$        & $1.99\pm0.29$      \\[0.1ex]
Q~$1327-206$   & $0.0174$  & $3.41\pm0.69$        & $3.39\pm0.64^{a}$  \\[0.1ex]
PG~$1427+480$  & $0.1203$  & $0.78\pm0.11$        & $0.66\pm0.11$      \\[0.9ex]\hline
\multicolumn{4}{l}{$^{a}$ Blended with Galactic {\MgI} $\lambda 2852$}
\end{tabular}
\end{center}
\end{table}

\begin{table}[p]
\begin{center}
\begin{tabular}{rc}
\multicolumn{2}{c}{TABLE 2. {\sc Equivalent Widths}}  \\[0.3ex]\hline\hline
   ID          & $W_{r}$, {\AA}       \\[0.7ex]\hline
\multicolumn{2}{c}{3C~232~~~$z=0.0050$} \\[0.7ex]\hline
{\FeII} 2344   & $1.75\pm 0.11$ \\
{\FeII} 2374   & $0.74\pm 0.11$ \\
{\FeII} 2382   & $2.01\pm 0.10$ \\
{\FeII} 2600   & $1.87\pm 0.09$ \\
{\MgII} 2796   & $2.54\pm 0.09$ \\
{\MgII} 2803   & $2.45\pm 0.10$ \\
{\ MgI} 2852   & $0.82\pm 0.13$ \\[0.7ex]\hline
\multicolumn{2}{c}{PKS~$0439-433$~~~$z=0.1012$} \\[0.7ex]\hline
{\FeII} 2344   & $1.18\pm 0.11$ \\
{\FeII} 2374   & $0.27\pm 0.10$ \\
{\FeII} 2382   & $1.27\pm 0.09$ \\
{\FeII} 2586   & $0.91\pm 0.10$ \\
{\FeII} 2600   & $1.23\pm 0.18$ \\
{\FeII} 2600   & $0.40\pm 0.20$ \\
{\MgII} 2796   & $1.43\pm 0.22$ \\
{\MgII} 2796   & $0.84\pm 0.24$ \\
{\MgII} 2803   & $1.19\pm 0.20$ \\
{\MgII} 2803   & $0.75\pm 0.23$ \\
{\MgI} 2852    & $0.54\pm 0.11$ \\[0.7ex]\hline
\multicolumn{2}{c}{Q~$1327-206$~~~$z=0.0174$} \\[0.7ex]\hline
{\FeII} 2344   & $1.45\pm 0.09^{a}$ \\
{\FeII} 2374   & $2.12\pm 0.12^{a}$ \\
{\FeII} 2382   & $2.12\pm 0.07^{a}$ \\
{\MgII} 2796   & $2.77\pm 0.32$ \\
{\MgII} 2796   & $0.65\pm 0.33$ \\
{\MgII} 2803   & $1.72\pm 0.38^{b}$ \\
{\MgII} 2803   & $1.66\pm 0.47^{b}$ \\
{\MgI} 2852    & $1.25\pm 0.06$ \\[0.7ex]\hline
\multicolumn{2}{c}{PG~$1427+480$~~~$z=0.1203$} \\[0.7ex]\hline
{\Lya}  1215   & $3.95\pm 0.26$ \\
{\CII}  1334   & $0.25\pm 0.03$ \\
{\FeII} 2344   & $0.36\pm 0.10$ \\
{\FeII} 2374   & $0.23\pm 0.10$ \\
{\FeII} 2382   & $0.47\pm 0.10$ \\ 
{\FeII} 2600   & $0.40\pm 0.10$ \\
{\MgII} 2796   & $0.77\pm 0.11$ \\
{\MgII} 2803   & $0.64\pm 0.11$ \\[0.7ex]\hline
\multicolumn{2}{l}{$^{a}$ In the {\Lya} forest} \\
\multicolumn{2}{l}{$^{b}$ Blended with Galactic {\MgI} $\lambda 2852$}
\end{tabular}
\end{center}
\end{table}

\begin{table}[p]
\begin{center}
\begin{tabular}{rccc}
\multicolumn{4}{c}{TABLE 3. {\sc Redshift Number Density at $z=0.05$}} \\[0.3ex]\hline\hline
$W_{min}$ & $\gamma$      & SS92                   & This Survey    \\[0.5ex]\hline
0.3~{\AA} & $0.78\pm0.42$ & $0.56^{+0.19}_{-0.14}$ & $< 0.32$               \\
0.6~{\AA} & $1.02\pm0.53$ & $0.25^{+0.11}_{-0.08}$ & $0.22^{+0.12}_{-0.09}$ \\
1.0~{\AA} & $2.24\pm0.76$ & $0.05^{+0.04}_{-0.02}$ & $0.16^{+0.09}_{-0.05}$ \\[0.3ex]\hline
\end{tabular}
\end{center}
\end{table}

\clearpage

\end{document}